# Quantum interference patterns around magnetic impurities in a 2D electron gas with strong Rashba spin-orbit interaction

N.V. Khotkevych[a], Yu.A. Kolesnichenko[a*], J.M. van Ruitenbeek[b]

[a] *B. Verkin Institute for Low Temperature Physics and Engineering, National Academy of Sciences of Ukraine, 47, Lenin Ave., 61103, Kharkov, Ukraine*

[b] *Huygens-Kamerlingh Onnes Laboratorium, Universiteit Leiden, Postbus 9504, 2300 Leiden, The Netherlands*

A B S T R A C T

Explicit dependencies of the local density of states and the magnetization local density of states have been obtained around magnetic point defects in a two-dimensional electron gas with Rashba spin-orbit interaction. The expressions are given in the Born approximation for arbitrary magnitude and orientation of the defect magnetic moment, and for arbitrary size of the constant of spin-orbit interaction $\alpha$. On the basis of our asymptotically exact formulas a procedure for the determination of the constant $\alpha$ in STM experiments is proposed. The novel analytical results are compared with the numerical and approximate results of previous work. At variance with earlier work we find that that magnetic scattering in the presence of spin-orbit interaction does not modify the local density of states and that the spin texture resulting from in-plain orientation of the defect magnetic moment, and elastic scattering, is not a purely in-plane spin texture.

## I. INTRODUCTION

Spin-orbit interaction (SOI) in a two-dimensional electron gas (2DEG) has attracted great current interest due to its key role for various newly discovered phenomena [1,2] and possible applications in spintronics [3]. One of the practical realizations of a 2DEG with SOI is exploiting surface states in metals [4]. By reason of the high parallel conductivity of the bulk of the metal the surface states cannot be observed in galvanomagnetic measurements, and surface sensitive techniques should be used. Accordingly, the spin-orbit splitting of surface states near (111) surfaces of noble metals has been found by angular-resolved photoemission spectroscopy with a high-momentum resolution [5,6,7].

One of the most powerful methods for the study of conducting surfaces is scanning tunneling microscopy (STM) [8]. In the framework of a simplified theoretical model of STM in Ref. [9] it was shown that the tunnel conductance measured by STM is proportional to the local density of surface states of the sample. In spite of a variety of assumptions, the theory [9] in most cases describes experiments adequately. The presence of isolated defects on flat metal surfaces opens interesting possibilities. Images obtained using STM display standing waves related to electron scattering by surface steps and single defects [10]. The physical origin of these patterns in the constant-current mode STM images is the same type of interference that leads to Friedel oscillations in the electron local density of states in the vicinity of a scatterer. The patterns arise as a result of quantum interference between incident electron waves and waves scattered by the defects. The analysis of STM images around defects provides information on the Fermi surface contours, notably for two-dimensional (2D) surface states [11,12,13].

---

[*] Corresponding author.
E-mail address: kolesnichenko@ilt.kharkov.ua (Yu.A. Kolesnichenko)

Magnetic structures also can be studied in real space on the nanometer scale using spin-polarized scanning tunneling microscopy (SP-STM), which provides real-space images of magnetic order with atomic resolution [14,15]. It was shown that the SP-STM current can be decomposed into a non-spin-polarized part, which depends on sample local density of states (LDOS), and a spin-polarized part. The spin-polarized part is proportional to the scalar product of the tip magnetization density of states and the vector of the magnetization local density of states (MLDOS) of the sample [16]. The local magnetic defect produces a long-range spin polarization of the conduction electrons, which oscillates with the distance from the defect [17]. These oscillations are often called RKKY polarization and they are the magnetic analogue of Friedel oscillations. In ref. [18] spin-polarized scanning tunneling spectroscopy has been used to reveal how the standing wave patterns of confined surface state electrons on top of nanometer-scale ferromagnetic Co islands on Cu(111) are affected by the spin character of the responsible state.

A number of recent experimental and theoretical works deals with the problem of quantum interference of 2D electrons around single point-like defect in the presence of SOI (see papers [19,20,21,22,23,24,25,26,27] an references therein). Particularly, it was noted that the SOI induced splitting of the surface states is unobservable in the surface charge density as probed with STM [22]. In the latter paper simple scattering approach was used (assuming elastic scattering) to demonstrate that for Rashba SOI the period of the standing waves is defined by the sum of the Fermi wave vectors for both contours of the split Fermi surface, because for a nonmagnetic scatterer spin conservation allows only backscattering with umklapp process between Fermi contours [22]. In Ref. [20] STM has been applied for spectroscopic maps of quasiparticle interference patterns around a single magnetic MnPc molecule on a surface with strong SOC: Bi(110). It was noted that the forbidden scattering channels are activated by a magnetic scattering center, but a fingerprint of the backscattering processes appears in the magnetization patterns, and not in the charge density, suggesting that only SP-STM can access this information [20]. The last statement is in contradiction with the result of ref. [24], in which it is shown that the correction to the electron density of states due to the local spin is in first order proportional to the exchange coupling and the strength of the SOI. According to Fransson [26] elastic scattering of a 2DEG with Rashba SOI gives rise to a purely in-plane spin texture for an in-plane magnetic scattering potential, out-of-plane components emerge upon activation of inelastic scattering processes.

From our point of view the relation between different conclusions obtained by different methods is not completely clear and further theoretical studies is needed. It is useful to express the LDOS and the MLDOS in explicit form (albeit in the framework of a simplified model). The analytical formulas will display in obvious form the dependencies of the harmonics in the interference patterns on the distance from the defect, the direction of the defect magnetic moment, the parameters of electron energy spectrum and the strength of the SOI.

In this paper we investigate the LDOS and the MLDOS around a magnetic point defect in a 2DEG with Rashba SOI. By a single set of calculations we study the manifestation of electron interference in the LDOS and the MLDOS, and their dependencies on the physical parameters of the system. In Sec.II we present the methods of calculations and find the Green function for 2D electrons scattered by a defect, in Born's approximation. The scalar scattering potential of the defect and its magnetic moment are taken into account simultaneously. Analytical formulas for the LDOS and the MLDOS at arbitrary value of the SOI constant are derived. In Sec.III the asymptotes of the LDOS and the MLDOS at large distances from the defect are obtained. The asymptotically exact formulas give the main harmonics in the oscillatory parts of the LDOS and the MLDOS and their dependencies on the distance from the defect, on the constant of Rashba SOI and on the direction of defect magnetic moment. In Sec. IV we conclude by discussing the physical interpretation of our results and the possibilities for their observation in STM experiments.

## II. MODEL AND BASIC EQUATIONS

Let us consider a free electron gas with Rashba SOI [28] confined to the $xy$ plane. A single magnetic defect with short-range scattering potential, treated as having a classical spin, is placed at the point $\mathbf{r} = 0$. We assume a spin of the defect $S \geq 1$ and neglect the Kondo screening of it, which has been treated in Ref. [19]. The problem is solved for a quadratic isotropic dispersion law of the 2D electrons. All results are found to first order in the electron interaction with the defect (Born's approximation).

In framework of the model described above the Hamiltonian $\hat{H}$ of the system can be written as

$$\hat{H} = \hat{H}_0 + \hat{V}(\mathbf{r}), \tag{1}$$

where $\hat{H}_0$ is the Hamiltonian of free-space 2DEG with Rashba SOI,

$$\hat{H} = \frac{\hbar^2 \hat{\mathbf{\kappa}}^2}{2m^*} \hat{\sigma}_0 + \alpha \left( \hat{\sigma}_x \hat{\kappa}_y - \hat{\sigma}_y \hat{\kappa}_x \right), \tag{2}$$

$\hbar \hat{\mathbf{\kappa}} = -i\hbar \nabla_\rho$ is the momentum operator, $m^*$ is the electron effective mass, $\alpha$ is the Rashba spin–orbit coupling strength, $\hat{V}(\mathbf{r})$ describes the electron interaction with the defect

$$\hat{V}(\mathbf{r}) = \left( \gamma \hat{\sigma}_0 + \mathbf{J}\hat{\mathbf{\sigma}} \right) \delta(\mathbf{r}) . \tag{3}$$

Here, $\gamma$ is constant of potential interaction of the electrons with the defect, $\mathbf{J}$ is the magnetic moment of the defect, $\hat{\mathbf{\sigma}} = (\hat{\sigma}_x, \hat{\sigma}_y, \hat{\sigma}_z)$ is the Pauli spin vector, $\hat{\sigma}_0$ is the $2 \times 2$ unit matrix. We consider a static magnetic moment assuming that the magnetization vector $\mathbf{J} = (J_x, J_y, 0)$ has only in-plane components. Note that a more realistic model of a short-range potential with a finite radius of action $r_V$ for the function $\hat{V}(\mathbf{r})$ gives the same results as the $\delta$ - function potential in the limit $\kappa r_V \ll 1$ ($\kappa$ is the electron wave vector). If $\kappa r_V \geq 1$, at the distances from the defect $|\mathbf{r}| \gg r_V$, $\kappa r \gg 1$ the finite value of $r_V$ manifests itself as a coefficient less than unity in the oscillatory part of the STM conductance [29].

In absence of the defect $\hat{V} = 0$ eigenvalues and eigenfunctions of the Hamiltonian (1) are well known [1]

$$\varepsilon_{1,2} = \frac{\hbar^2 \kappa^2}{2m^*} \pm \alpha \hbar \kappa > 0, \tag{4}$$

$$\hat{\psi}_{1,2}(\mathbf{r}) = \frac{1}{\sqrt{2}} e^{i\mathbf{\kappa}\mathbf{r}} \begin{pmatrix} 1 \\ \pm i e^{i\phi} \end{pmatrix}, \tag{5}$$

where the phase $\phi$ represents the angle between $\mathbf{\kappa}$ and the $y$ axis. The spin orientation axis $\mathbf{n}_{1,2}$ is given by the expectation value

$$\mathbf{n}_{1,2} = \left\langle \psi_{1,2}^{(0)} \middle| \mathbf{\sigma} \middle| \psi_{1,2}^{(0)} \right\rangle = \mp (\sin\theta, -\cos\theta, 0). \tag{6}$$

It is perpendicular to the wave vector $\mathbf{n}_{1,2} \perp \mathbf{\kappa} = \kappa(\cos\theta, \sin\theta, 0)$.

As follows from Eqs. (4) - (6), the opposite spins have different energies and the Fermi surface splits into two concentric circles

$$\varepsilon_{1,2}(\kappa) = \varepsilon_F. \tag{7}$$

The local density of states $\rho$ and the local magnetization density of states $\mathbf{M}$ at the Fermi level $\varepsilon = \varepsilon_F$ can be expressed in terms of the retarded, single-particle Green's function, $G^R(\mathbf{r},\mathbf{r},\varepsilon_F)$, as

$$\rho(\mathbf{r},\varepsilon_F) = -\frac{1}{\pi}\operatorname{Im} Tr_\sigma\left[\hat{G}^R(\mathbf{r},\mathbf{r},\varepsilon_F)\right], \qquad (8)$$

$$\mathbf{M}(\mathbf{r},\varepsilon_F) = -\frac{1}{\pi}\operatorname{Im} Tr_\sigma\left[\hat{\boldsymbol{\sigma}}\hat{G}^R(\mathbf{r},\mathbf{r},\varepsilon_F)\right], \qquad (9)$$

where $\mathbf{r}=(x,y)$ is the coordinate in the plane of the 2DEG. The real-space Green's function $G^R(\mathbf{r},\mathbf{r},\varepsilon_F)$ is calculated below to first approximation in the interaction potential (3). By using the simple relation $\hat{H}_{SO}^2 = \alpha^2 \kappa^2 \hat{\sigma}_0$ we can rewrite the Green function in zeroth approximation in the potential (3) in momentum representation as follows

$$\hat{g}_0(\boldsymbol{\kappa},\varepsilon) = \frac{1}{\varepsilon - \hat{H}_0} \equiv \frac{\left(\varepsilon - \frac{\hbar^2\kappa^2}{2m^*}\right)\hat{\sigma}_0 + \hat{H}_{SO}}{\left(\varepsilon - \frac{\hbar^2\kappa^2}{2m^*}\right)^2 - \alpha^2\kappa^2}. \qquad (10)$$

After algebraic transformations Eq.(10) can be presented in the form [30]

$$\hat{g}_0(\boldsymbol{\kappa},\varepsilon) = \frac{m^*}{\tilde{\kappa}\hbar^2}\left[\frac{\kappa_1\hat{\sigma}_0 - (\hat{\sigma}_x\hat{\kappa}_y - \hat{\sigma}_y\hat{\kappa}_x)}{\kappa_1^2 - \kappa^2} + \frac{\kappa_2\hat{\sigma}_0 + (\hat{\sigma}_x\hat{\kappa}_y - \hat{\sigma}_y\hat{\kappa}_x)}{\kappa_2^2 - \kappa^2}\right]. \qquad (11)$$

In Eq.(11)

$$\tilde{\kappa} = \sqrt{\frac{2m^*\varepsilon}{\hbar^2} + \left(\frac{m^*\alpha}{\hbar^2}\right)^2}; \qquad (12)$$

$$\kappa_{1,2} = \tilde{\kappa} \mp \frac{m^*\alpha}{\hbar^2}. \qquad (13)$$

By using Eq. (11) one can find the retarded Green function in the real-space representation

$$\hat{G}_0^R(\mathbf{r}-\mathbf{r}',\varepsilon) = \int\frac{d\boldsymbol{\kappa}}{(2\pi)^2}\hat{g}_0(\boldsymbol{\kappa},\varepsilon+i0)e^{i\boldsymbol{\kappa}(\mathbf{r}-\mathbf{r}')}. \qquad (14)$$

The integrals in Eq. (14) can be expressed in terms of the Hankel functions. The integral with $\kappa_{1,2}\hat{\sigma}_0$ is known as the Green function of a 2D wave equation [31]

$$\int\frac{d\boldsymbol{\kappa}}{(2\pi)^2}\frac{e^{i\boldsymbol{\kappa}(\mathbf{r}-\mathbf{r}')}}{\kappa_{1,2}^2 - \kappa^2 + i0} = -\frac{i}{4}H_0^{(1)}(\kappa_{1,2}|\mathbf{r}-\mathbf{r}'|). \qquad (15)$$

It is easy to note that integrals with $\kappa_i\hat{\sigma}_k$ can be found as

$$\int \frac{d\kappa}{(2\pi)^2} \frac{\kappa_i e^{i\kappa(\mathbf{r}-\mathbf{r}')}}{\kappa_{1,2}^2 - \kappa^2 + i0} = \frac{\partial}{i\partial r_i} \int \frac{d\kappa}{(2\pi)^2} \frac{e^{i\kappa(\mathbf{r}-\mathbf{r}')}}{\kappa_{1,2}^2 - \kappa^2 + i0} = \frac{\kappa_{1,2}(r_i - r_i')}{4|\mathbf{r}-\mathbf{r}'|} H_1^{(1)}(\kappa_{1,2}|\mathbf{r}-\mathbf{r}'|). \quad (16)$$

Finally, using equations (11)-(16) we find

$$\hat{G}_0^R(\mathbf{r},\mathbf{r}';\varepsilon) = \frac{im^*}{4\tilde{\kappa}\hbar^2}\left\{\left(\kappa_1 H_0^{(1)}(\kappa_1|\mathbf{r}-\mathbf{r}'|) + \kappa_2 H_0^{(1)}(\kappa_2|\mathbf{r}-\mathbf{r}'|)\right)\hat{\sigma}_0 - i\hbar\frac{\left(\hat{\sigma}_y(x-x') - \hat{\sigma}_x(y-y')\right)}{|\mathbf{r}-\mathbf{r}'|}\left(\kappa_1 H_1^{(1)}(\kappa_1|\mathbf{r}-\mathbf{r}'|) - \kappa_2 H_1^{(1)}(\kappa_2|\mathbf{r}-\mathbf{r}'|)\right)\right\}. \quad (17)$$

Substituting the Green function of zeroth approximation (17) in the Eqs. (8), (9), one can verify that we recover the density of states of a free 2DEG $\rho_0 = m^*/\pi\hbar^2$ and zero magnetization $\mathbf{M}_0 = 0$.

The electron Green's function $\hat{G}^R(\mathbf{r},\mathbf{r}',\varepsilon)$ including the effect of a scattering center in linear approximation in $\hat{V}(\mathbf{r})$ is written as

$$\hat{G}^R(\mathbf{r},\mathbf{r}',\varepsilon) = \hat{G}_0^R(\mathbf{r},\mathbf{r}';\varepsilon) + \hat{G}_0^R(\mathbf{r},0;\varepsilon)(\gamma\hat{\sigma}_0 + \mathbf{J}\hat{\boldsymbol{\sigma}})\hat{G}_0^R(0,\mathbf{r}';\varepsilon); \quad \mathbf{r},\mathbf{r}' \neq 0. \quad (18)$$

Eqs. (17), (18) provide the possibility to find analytical formulas for the LDOS (8) and the MLDOS (9) in the Born approximation for arbitrary constant of SOI $\alpha$.

**II. LOCAL DENSITIES OF STATES**

By using the Eqs. (8), (9) (17), (18) we can find LDOS $\rho$ and MLDOS $\mathbf{M}$. All formulas below have been obtained without any additional assumptions. After cumbersome calculations one obtains,

$$\rho(\mathbf{r},\varepsilon_F) = \frac{m^*}{\pi\hbar^2}\left\{1 + \frac{m^*\gamma}{4\hbar^2\tilde{\kappa}^2}[(\kappa_1 J_0(\kappa_1 r) + \kappa_2 J_0(\kappa_2 r))(\kappa_1 Y_0(\kappa_1 r) + \kappa_2 Y_0(\kappa_2 r))\right.$$
$$\left. + (\kappa_1 J_1(\kappa_1 r) - \kappa_2 J_1(\kappa_2 r))(\kappa_1 Y_1(\kappa_1 r) - \kappa_2 Y_1(\kappa_2 r))]\right\}; \quad (19)$$

$$M_x(\mathbf{r},\varepsilon_F) = \frac{m^{*2}}{4\pi\hbar^4\tilde{\kappa}^2 r^2}\left\{J_x r^2(\kappa_1 J_0(\kappa_1 r) + \kappa_2 J_0(\kappa_2 r)) \times (\kappa_1 Y_0(\kappa_1 r) + \kappa_2 Y_0(\kappa_2 r)) - (J_x(x^2 - y^2) + 2J_y xy)\right.$$
$$\left. (\kappa_1 J_1(\kappa_1 r) - \kappa_2 J_1(\kappa_2 r))(\kappa_1 Y_1(\kappa_1 r) - \kappa_2 Y_1(\kappa_2 r))\right\}; \quad (20)$$

$$M_y(\mathbf{r},\varepsilon_F) = \frac{m^{*2}}{4\pi\hbar^4\tilde{\kappa}^2 r^2}\left\{J_y r^2(\kappa_1 J_0(\kappa_1 r) + \kappa_2 J_0(\kappa_2 r)) \times (\kappa_1 Y_0(\kappa_1 r) + \kappa_2 Y_0(\kappa_2 r)) + (J_y(x^2 - y^2) - 2J_x xy)\right.$$
$$\left. (\kappa_1 J_1(\kappa_1 r) - \kappa_2 J_1(\kappa_2 r))(\kappa_1 Y_1(\kappa_1 r) - \kappa_2 Y_1(\kappa_2 r))\right\}; \quad (21)$$

$$M_z(\mathbf{r},\varepsilon_F) = \frac{m^{*2}}{4\pi\hbar^4\tilde{\kappa}^2 r}\{(J_x x + J_y y)[(\kappa_1 J_1(\kappa_1 r) - \kappa_2 J_1(\kappa_1 r))(\kappa_2 Y_0(\kappa_1 r) + \kappa_2 Y_0(\kappa_2 r)) \quad (22)$$
$$+ (\kappa_1 J_0(\kappa_1 r) + \kappa_2 J_0(\kappa_2 r))(\kappa_1 Y_1(\kappa_1 r) - \kappa_2 Y_1(\kappa_2 r))]\},$$

where $r = \sqrt{x^2 + y^2}$. Here and below wave vectors $\kappa_{1,2}$ (13) and $\tilde{\kappa}$ (12) are taken at the Fermi energy $\varepsilon = \varepsilon_F$, $J_k(x)$ and $Y_k(x)$ are Bessel functions of the first and second kind, and of order $k$. By reason of the divergences of functions $Y_{0,1}(x)$ at $x \to 0$ (see, for example [32]), the equations (19) - (22) are correct for distances from the defect $\kappa_{1,2} r \geq 1$, for which the corrections obtained in the framework of the perturbation procedure in terms of the strength of the scattering potential (3) are small. At $\alpha = 0$, when the SOI is absent, and Eq. (19) transforms into the known result for oscillations of the LDOS for a free 2DEG [33], and $M_{x,y} \sim J_{x,y}$, $M_z = 0$.

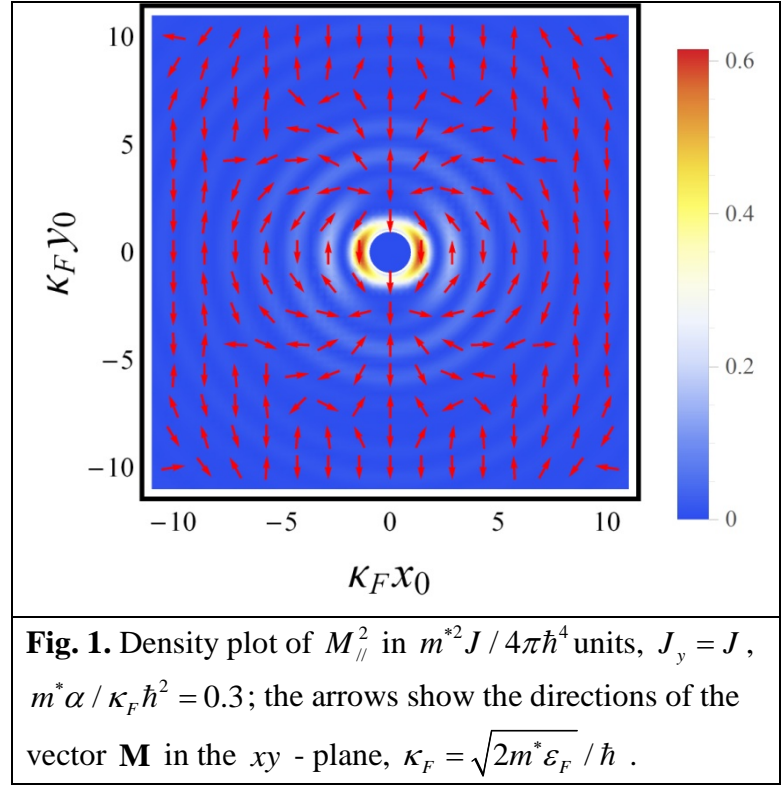

**Fig. 1.** Density plot of $M_{//}^2$ in $m^{*2}J/4\pi\hbar^4$ units, $J_y = J$, $m^*\alpha/\kappa_F \hbar^2 = 0.3$; the arrows show the directions of the vector **M** in the $xy$-plane, $\kappa_F = \sqrt{2m^*\varepsilon_F}/\hbar$.

In Fig. 1 we present a density plot of $M_{//}^2 = |M_x|^2 + |M_y|^2$ for finite SOI, plotted by using Eqs. (20), (21), where the arrows show the directions of the vector **M** in the $xy$-plane, for the magnetic moment **J** oriented along the $y$ direction. A contour plot of the function (22) shown in Fig. 2 illustrates the distribution of the $M_z$ component of MLDOS in the $xy$-plane. In Fig. 1 and Fig. 2 we remove the region $\kappa_F r < 1$ where our formulas are not valid.

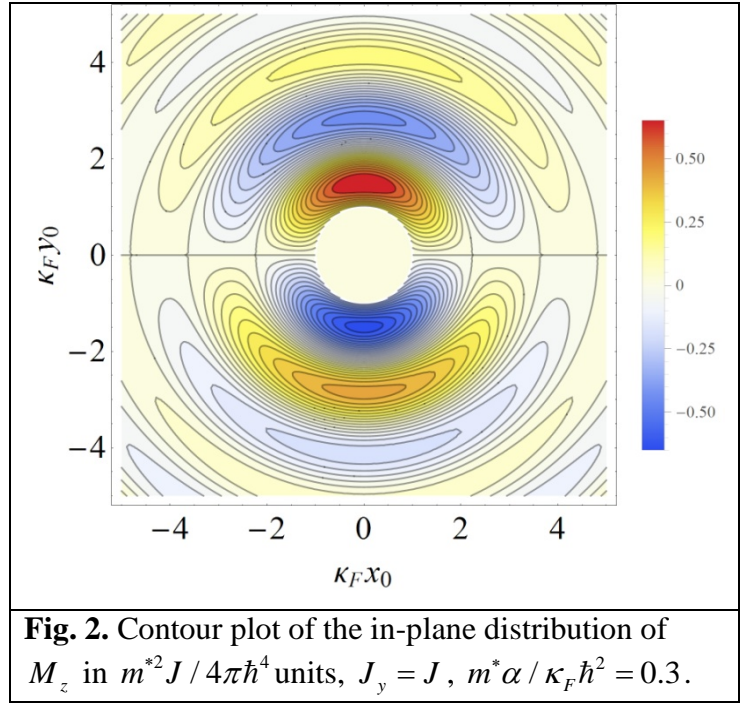

**Fig. 2.** Contour plot of the in-plane distribution of $M_z$ in $m^{*2}J/4\pi\hbar^4$ units, $J_y = J$, $m^*\alpha/\kappa_F\hbar^2 = 0.3$.

From Eqs. (19) - (22) one can draw the following conclusions: In the Born approximation the spatial oscillations of the LDOS are not affected by the magnetic scattering channels, as happens in the absence of SOI (see [17], p.363). The 2DEG is spin-polarized around the defect. The net polarization $p(\mathbf{r}, \varepsilon_F)$ is given by the difference of the diagonal elements $\hat{G}^R_{11(22)}(\mathbf{r}, \mathbf{r}, \varepsilon_F)$ of the Green's function matrix (17)

$$p(\mathbf{r},\varepsilon_F) = \rho_\uparrow(\mathbf{r},\varepsilon_F) - \rho_\downarrow(\mathbf{r},\varepsilon_F) = -\frac{1}{\pi}\mathrm{Im}\left[\hat{G}^R_{11}(\mathbf{r},\mathbf{r},\varepsilon_F) - \hat{G}^R_{22}(\mathbf{r},\mathbf{r},\varepsilon_F)\right] = M_z(\mathbf{r},\varepsilon_F) \quad (23)$$

The SOI makes the MLDOS oscillations anisotropic, i.e. their amplitude depends on the direction of the magnetic moment of the defect, $\mathbf{J}$. For the in-plain orientation of the vector $\mathbf{J}$, the $z$ component of magnetization $M_z(\mathbf{r}, \varepsilon_F)$ (22) obtains a finite value.

## II. ASYMPTOTES OF LDOS AND MLDOS AT LARGE DISTANCES FROM THE DEFECT

For a clear interpretation of the obtained results (19) - (22) we consider the asymptotic behavior of the LDOS and the MLDOS at large distances from the defect $\kappa_{1,2} r \gg 1$. The corresponding asymptotic formulas can be easily obtained by using the asymptotes of Bessel functions for large arguments [32]. For the LDOS we find

$$\rho(r, \varepsilon_F) = \frac{m^*}{\pi\hbar^2}\left[1 - \frac{m^* g}{\pi\hbar^2 \tilde{\kappa}^2 r}\sqrt{\kappa_1 \kappa_2} \cos\left((\kappa_1 + \kappa_2)r\right)\right], \quad (24)$$

where

$$\kappa_1(\varepsilon_F)\kappa_2(\varepsilon_F) = \frac{2m^*\varepsilon_F}{\hbar^2}; \quad \kappa_1(\varepsilon_F) + \kappa_2(\varepsilon_F) = 2\tilde{\kappa}(\varepsilon_F) = 2\sqrt{\frac{2m^*\varepsilon_F}{\hbar^2} + \left(\frac{m^*\alpha}{\hbar^2}\right)^2}. \quad (25)$$

The physical origin of the oscillatory term in Eq. (24) is the quantum interference of electron waves incident to the defect with backscattered waves. At large distances from the defect the main contribution to the oscillatory part of 2D LDOS originates from small parts (both, for incident and backscattered electrons) of the Fermi contour near the points in which the velocity is collinear to the vector $\boldsymbol{\rho}_0$. For the split Fermi contours (7) these points can belong to the same or different contours. In Eq. (24) the period of oscillations $\Delta r_{12} = 2\pi/(\kappa_1 + \kappa_2)$ is defined by the radii $\kappa_{1,2}(\varepsilon_F)$ of both Fermi contours $\varepsilon_{1,2} = \varepsilon_F$. It means that interference takes place between states belonging to different contours but having the same spin. Such result is in accordance with conclusions of Ref. [22].

Under the condition $\kappa_{1,2}r \gg 1$ formulas (20), (21), (22) for the components of the MLDOS are simplified to the following form

$$M_x(\mathbf{r},\varepsilon_F) = \frac{m^{*2}}{2\pi^2\hbar^4\tilde{\kappa}^2 r}\Big[2\mathbf{Jn}_\perp \sin\theta\sqrt{\kappa_1\kappa_2}\cos((\kappa_1+\kappa_2)r) \\ +\mathbf{Jn}_\parallel \cos\theta(\kappa_1\cos(2\kappa_1 r) + \kappa_2\cos(2\kappa_2 r))\Big], \quad (26)$$

$$M_y(\mathbf{r},\varepsilon_F) = \frac{m^{*2}}{2\pi^2\hbar^4\tilde{\kappa}^2 r}\Big[-2\mathbf{Jn}_\perp \cos\theta\sqrt{\kappa_1\kappa_2}\cos((\kappa_1+\kappa_2)r) \\ +\mathbf{Jn}_\parallel \sin\theta(\kappa_1\cos(2\kappa_1 r) + \kappa_2\cos(2\kappa_2 r))\Big], \quad (27)$$

$$M_z(\mathbf{r},\varepsilon_F) = \frac{m^{*2}}{2\pi^2\hbar^4\tilde{\kappa}^2 r}\mathbf{Jn}_\parallel(\kappa_1\sin(2\kappa_1 r) - \kappa_2\sin(2\kappa_2 r)). \quad (28)$$

Here we introduced two unit vectors $\mathbf{n}_\parallel \parallel \mathbf{r}$ and $\mathbf{n}_\perp \perp \mathbf{r}$

$$\mathbf{n}_\parallel = (\cos\theta,\sin\theta,0); \quad \mathbf{n}_\perp = (\sin\theta,-\cos\theta,0), \quad (29)$$

$\theta$ is the angle between vector $\mathbf{r}$ and $x_0$ axis, $\mathbf{r} = r\mathbf{n}_\parallel$. Comparing Eqs. (29) and (6) one can see that the scalar products $\mathbf{Jn}_\parallel$ and $\mathbf{Jn}_\perp$ are the projections $\mathbf{J}$ the perpendicular and parallel to spin directions, i.e the projections of vector $\mathbf{J}$ parallel and perpendicular to the vector $\mathbf{r}$. Terms with $\mathbf{Jn}_\parallel$ describe the contribution to the LSDOS of spin-flip processes, for which the backscattered state belongs to the same Fermi contour, while terms with $\mathbf{Jn}_\perp$ correspond to scattering events with spin conservation, as applies for potential scattering (24). Such differences in electron scattering by magnetic impurities in the presence of SOC results in anisotropic magnetoresistance of diluted magnetic semiconductors [34].

## III. CONCLUSIONS

The backscattering by a non-magnetic defect in a 2DEG with Rashba SOC is not suppressed, but the condition of spin conservation requires that the scattering process must be accompanied by a transition between the two branches of the energy spectrum (transitions between states $2\leftrightarrow 4$ and $1\leftrightarrow 3$ in figure 3). The period of the oscillations in the LDOS is defined by the arithmetic mean of the radii of the Fermi contours

$$\Delta r_{12} = \pi / \tilde{\kappa}(\varepsilon_F), \tag{30}$$

where $\tilde{\kappa}(\varepsilon_F)$ is given by Eq. (25) and depends on the constant of SOC $\alpha$. The amplitude of the LDOS oscillations (24) is proportional to the constant $\gamma$ of potential interaction of the electrons with the defect and is of the same order of magnitude as for a free 2DEG [33] (compare with result of Ref.[24]). Generally, the constant $\alpha$ can be found from the period (30), if the Fermi energy is known.

The magnetic moment of the defect locally breaks the time-reversal symmetry, which opens scattering channels assisted by spin-flip processes. As was shown in Ref. [34] for a 2DEG with Rashba SOC, when the electron spin is perpendicular to the magnetization vector of the defect, i.e. when the electron moves along the vector $\mathbf{J}$, the backscattering occurs due to transitions between states on the same Fermi contour (transitions between states $1\leftrightarrow 4$ and $2\leftrightarrow 3$ in Fig. 3), while backscattering for electrons moving a the direction perpendicular to the vector $\mathbf{J}$ results in transitions between energy branches, similar as for scattering on a scalar potential. These facts are reflected in the formulas for the components of the MLDOS (26) - (28), in which terms proportional to $\mathbf{Jn}_\perp$ oscillate with period (30) and terms proportional to $\mathbf{Jn}_\parallel$ have two harmonics with different periods of oscillations

$$\Delta r_{1(2)} = \pi / \kappa_{1(2)}(\varepsilon_F), \tag{31}$$

where $\kappa_{1(2)}(\varepsilon_F)$ are the radii (12) of Fermi contours (7). Formula (31) makes it possible to find the strength of the Rashba SOC easily

$$\frac{1}{\Delta r_2} - \frac{1}{\Delta r_1} = \frac{2m^*\alpha}{\pi\hbar^2}. \tag{32}$$

The amplitude of the spin-flip assisted oscillation is maximal when $\mathbf{Jn}_\perp = 0$, i.e. in the direction of magnetic moment of the defect.

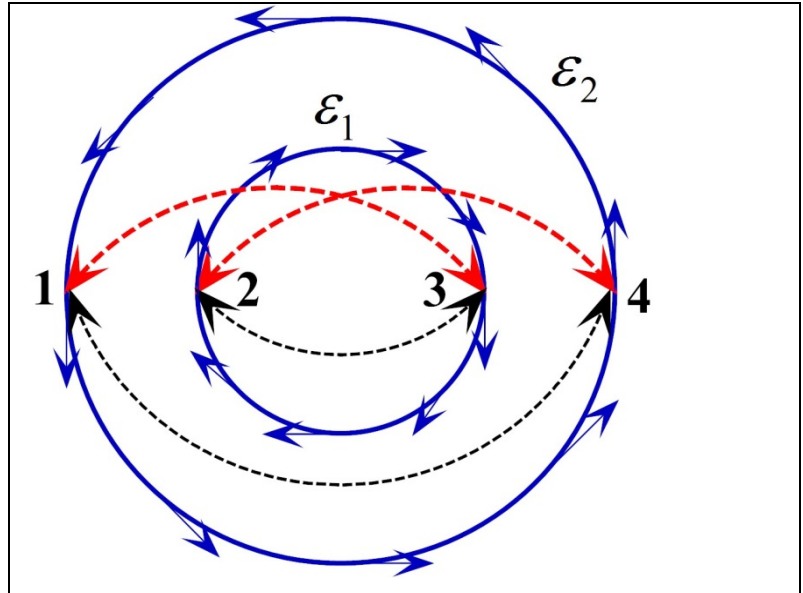

**Fig. 3.** Schematic illustration of possible transitions between states on Fermi contours (7) with opposite directions of the electron wave vectors. Arrows show the spin directions in the corresponding points of the Fermi contours $\varepsilon_{1,2} = \varepsilon_F$. Red arrows illustrate scattering events on non-magnetic defects or on a magnetic moment parallel to electron spin, black arrows connect states for scattering on a magnetic moment perpendicular to the electron spin.

Equation (28) shows that the magnetic moment of a defect situated in the plane of the 2DEG in the presence of SOI produces a non-zero magnetization direction perpendicular to the plane for the elastic electron scattering as well as for non-elastic scattering [26]. The periods of the oscillations of the $M_z$ component of the MLDOS show that this magnetization is related to the spin-flip scattering processes. As opposed to the components $M_{x,y}$ the magnetization along the $z$ axis is determined by the difference of the contributions from different Fermi contours.

Thus, first in the framework of the Born approximation explicit formulas for the LDOS and the MLDOS around a magnetic point defect in a 2DEG with an arbitrary constant Rashba spin-orbit interaction are obtained. We show that magnetic channels of the electron scattering do not contribute to the LDOS as it occurs in the absence of SOI. This conclusion refutes the existence of terms proportional to the defect magnetic moment in the LDOS, as found in the Ref. [24]. We find that the oscillatory dependence of MLDOS on the distance from the defect is the superposition of three harmonics. One of them has the period which depends on the sum of the radii of the different Fermi contours split by the SOI and two others depends on the doubled radius for each Fermi contour that makes it possible to find the constant of Rashba SOI easily. Our novel result is the finding the dependences of amplitudes of oscillations on the direction of the defect magnetic moment **J** in the plane of the 2DEG. The appearance of a component of the magnetization perpendicular to the plane for in-plane orientation of vector **J** as the result of elastic spin-flip processes in the presence of SOI is predicted. This contradicts the conclusion in Ref. [26] that elastic scattering gives rise to a purely in-plane spin texture for an in-plane magnetic scattering potential.